\documentclass[12pt]{iopart}

\usepackage{iopams}
\usepackage{graphicx}
\begin{document}
\title{Magnon-Mediated Pairing and Isotope Effect in Iron-based Superconductors}
\author{Jiansheng Wu$^1$ and Philip Phillips$^2$}
\address{$^1$ Department of Physics and Astronomy, University of California, Irvine, Irvine CA 92697-4575 , U.S.A. }
\ead{jianshew@uci.edu}

%\author{Philip Phillips}
\address{$^2$ Department of Physics, University of Illinois at
Urbana-Champaign, 1110 W. Green Street, Urbana IL 61801, U.S.A.}

\begin{abstract}
Within a minimal model for the iron-based superconductors in which itinerant electrons interact with a band of local moments, we derive a
%criterion for sign-reversing superconductivity.
a general conclusion for multi-band superconductivity. 
%=======================================
In a multi-band superconductor, due to the Adler theorem, the inter-band scattering dominates the intra-band scattering at the 
long wave length limit as long as both interactions are induced by 
Goldstone boson (which is magnon in our case) and the transfered momentum is nonzero. Such kind of interaction leads to a well-known sigh-reversing superconductivity even
 if the inter-band and intra-band interaction are repulsive.
%=======================================
This effect can be modeled as arising from an internal
Josephson link between the Fermi surface sheets. 
%=======================================
Our model is also consistent with the recently discovered 
%%%%%%%%%%%%%%%%%%%%%%%%%%%%
coexistence of superconductivity and magnetic order in iron-pnictides.
%=======================================
Although the experimentally observed isotope effect is large, $\alpha=0.4$, we show that it is
consistent with a non-phononic mechanism in which it is the isotope effects result in a change in the lattice constant and as a consequence the
zero-point motion of the Fe atoms.
\end{abstract}

%\submitto{\JPG}
\maketitle

\section{Introduction}
 Although the history of superconductivity in iron-based materials is
short\cite{JACS,Wang,Cruze}, a consensus seems to have been reached
that phonons are not the efficient cause of the pairing mechanism.
For example, density functional calculations\cite{Boeri} predict
that for LaFeAsO$_{1-x}$F$_x$, the dimensionless electron-phonon coupling
constant is $\lambda=0.21$ and the logarithmic-averaged frequency is
$206 \rm K$. In conjunction with the standard BCS expression for
$T_c$, these values lead to a
 transition temperature of $0.8\rm K$ which is
too low to explain $26\rm K$ superconductivity. Experimentally, the bare phonon
density of states \cite{PDOS} in LaFeAsO$_{1-x}$F$_x$  (hereafter 1111) is in
perfect agreement with the density functional results\cite{Boeri} and hence
corroborates that phonons are too weak to explain superconductivity. As a
result, mechanisms, such as spin fluctuations, which exploit the proximity of
the superconducting phase to the antiferromagnet with $(\pi,0)$
order\cite{Cruze} in the parent material have risen to the fore\cite{Singh,s}.

Consequently, it has come as a surprise that the iron-pnictdes have
a sizeable isotope effect\cite{isotope}. When $^{56}\rm Fe$ is
replaced by $^{54}\rm Fe$ in Ba$_{1-x}$K$_x$Fe$_2$As$_2$ (hereafter
122), the transition temperatures for the magnetic order at $x=0$ and
superconductivity at $x=0.4$ are changed by 1.39\% and 1.34\%,
respectively. Expressing the mass dependence of $T_c$ using $
\alpha=-d\ln T_c/d\ln M $, leads to a value of $\alpha=0.4$ for the
observed change in $T_c$. A full isotope effect in standard low
$T_c$ materials corresponds to $\alpha=0.5$\cite{Schrieffer}. Hence,
the pnictides at the doping levels studied have an isotope effect
comparable to that of elemental superconductors.   The origin of
this effect is not known but certainly presents a challenge for
purely electronic mechanisms of superconductivity in which $\alpha$
is supposedly negligible.

Another important aspect of iron-based superconductors is the competition and
coexistence of magnetic order and superconductivity. As the doping increases in
the 1111 materials, they undergo a sharp transition from a magnetically ordered
state to a superconducting one\cite{Cruze}. In contrast, both magnetism and
superconductivity coexist (though in distinct regions in the sample) in the 122
materials in a limited range of doping\cite{Goko}.  In $\rm SmFeAsO_{0.85}$,
these two order parameters coexist at a macroscopic level\cite{Coex}. In $\rm
SmFeAsO_{0.85}$, these two order parameters coexist at a macroscopic
level\cite{Coex}.
It was also reported that the superconductivity and magnetism
coexist in the Co-doped material BaFe$_{2-x}$Co$_x$As$_2$\cite{Coex1} as well
as Ba$_{1-x}$ K$_x$Fe$_2$As$_2$ and $Sr_{1-x}$K$_x$Fe$_2$As$_2$\cite{Coex2}.
One of the goals of this paper is to provide a microscopic
model to explain these phenomena.

%Moreover, $\rm Sm(O,F)FeAs$,$\rm (Ba,K)Fe_2As_2$ and $\rm A
%Fe_2As_2$,$\rm A=(Ca,Ba,Sr)$ show the coexistence of static
%magnetism with SC in a small concentration region around the phase
%boundary in a phase-separation way\cite{Goko}. In $\rm La(O,F)FeAs$
%and $\rm SmFeAsO_{0.85}$, AF and SC coexist in a macroscopic
%level\cite{Coex}. All these confirm that magnetic order compete and
%cooperate with SC. Based on these, spin-fluctuation might contribute
%to the pairing mechanism just as the heavy fermion SC.

In this paper, we address the apparent contradiction between the
irrelevance of phonons to Cooper pair-formation and the sizeable
isotope effect in the pnictides.  Aside from the isotope effect, any
pairing mechanism must resolve the high critical temperature and the
competition and between magnetism { ---} that is, the $(\pi,0)$
antiferromagnet, and superconductivity.   Within a two-band
model\cite{WPA} for the Fe-As layer in which local moments in one
band interact with itinerant electrons in the other, we show that
interband scattering mediated by magnons leads to a high critical
temperature. Our work provides a microscopic framework for understanding
 the well-known result\cite{Kondo,Singh} that
sign-reversing pairing does not require attractive inter nor
intra-band interactions in two-band models. Second, we show that
although the pairing in this
 model is electronic in nature, an isotope effect exists as a result of
 the sensitivity to the lattice constant.  A value of $\alpha=0.4$ requires a
 zero-point motion of the Fe ion to be $6.2 meV$, in agreement with
 experimentally known\cite{zeroFe} values.

 An important fact about the pnictides is that
 they exhibit $(\pi,0)$ antiferromagnetism but they are
 metallic, nonetheless.  Metallic behavior in the presence of such order
implies that some states at the chemical potential remain ungapped.
Whether the magnetism arises from local or weak-coupling physics is
currently not resolved\cite{hassan}.  However, models in both of
these extremes are problematic. For example, pure spin models such
as the $J_1$-$J_2$ model\cite{si, sachdev, jph} which rely on a
fine-tuning of the nearest and next-nearest neighbor exchange
interactions are clearly incomplete as they describe insulators. At
the other extreme, weak-coupling scenarios based on
nesting\cite{Dong,McGuire,Vild} are also problematic because the
monoclinic distortion\cite{Cruze, Vild} that precedes the
magnetically ordered phase in the pnictides shifts each half-filled
band away from the perfect nesting condition by an amount related to
the crystal field splitting. Regardless of which model is used, it
must at least account for the experimental fact that local magnetic
correlations exist above the ordering temperature as
evidenced\cite{local} by the temperature dependence of the
peak-to-peak linewidth and g factor in electron spin resonance
studies on LaFeAsO$_{1-x}$F$_x$. The simplest scenario that is
consistent with this physics is a two-orbital model in which an
itinerant band of electrons hybridizes with a set of local moments. The itinerant and
local moments reside on two orthogonal levels.
In a previous paper\cite{WPA},
we have shown that the multi-orbital physics of the Fe-As
interactions reduces to an effective two-orbital model.  The
itinerant-localized dichotomy arises from the difference in the p-d
character of the two levels. While the p-d hybridization in this
model appears as a tuning parameter, it was found\cite{WPA} that a
hybridization of 0.8eV was needed to explain the magnitude of the
magnetic moment. This value of theb hybridization is consistent with
the only available\cite{hybridization} experimental estimate.
 A further key prediction of this
work is that aside from the moment lying in the $a-b$ plane, a residual moment
lies along the $z-$axis with a magnitude of $0.06\mu_B$, as found
experimentally\cite{McGuire}. In addition, recent neutron scattering
 experiments\cite{neutron} have inferred from the spin-wave spectrum
  that the magnetism in CaFe$_2$As$_2$ is indeed caused by a complicated 
  mixture of localized and itinerant physics consistent with the model proposed recently\cite{WPA}.

\section{Magnon-Mediated Pairing}
\subsection{Itinerant-Localized Model}
In light of the success of this model in describing  the parent
magnetically ordered state, we adopt it here to investigate possible
superconducting instabilities. The model can be written as a
spin-fermion-like Hamiltonian\cite{magnon,Yamada},
\begin{eqnarray}
    H&=&H_{\rm e}+H_{\rm s}+H_{\rm sf}\\
    H_{\rm e}&=& \sum \epsilon_{{\bf k},\sigma}c^{\dag}_{{\bf k},\sigma}c_{{\bf k},\sigma}\nonumber\\
    H_{\rm s}&=&J_1\sum_{n.n}{\bf S_i}\cdot{\bf S_j}+J_2\sum_{n.n.n}{\bf S_i}\cdot{\bf S_j}\nonumber\\
    H_{\rm sf}&=&J \sum c^{\dag}_{{\bf i},\alpha}  {\vec \sigma}_{\alpha\beta}  c_{{\bf
    i},\beta}\cdot {\bf S_i}.
\end{eqnarray}
in which $H_{\rm e}$ describes the itinerant electrons, $H_{\rm s}$
the localized electrons which yield the magnetism and $H_{\rm sf}$
the spin-fermion interaction between the two sets of electrons. Here
$\epsilon_{\bf k}$ is the band structure which yields the Fermi
surface of the non-interacting system, $c^\dagger_{\bf k,\sigma}$
creates an electron with momentum $\bf k$  and spin $\sigma$, and
$S_i$ represents the spin on site $i$.  We have retained only the
nearest and next-nearest spin exchange interactions, $J_1$ and
$J_2$, respectively, as experiments and theory\cite{si,sachdev,DFA}
indicate that interactions beyond these are negligible. Although
$J_1$ and $J_2$ will be treated as phenomenological parameters, it is
important to note that they are both directly proportional to the
p-d hybridization. That is, if the p-d hybridization is ignored, as
advocated recently\cite{sawatzky}, both vanish.
The most important term is the Kondo coupling term $H_{\rm sf}$
where $J<0$.  A key assumption in this work then is that the magnetism
 is due to local moments
and not spin-density wave formation arising from a Fermi surface instability.  
As a result, in our work here, we do not retain the particle-hole instabilities 
which are of $O(J^2)$ but focus entirely on the particle-particle channel to 
develop a microscopic model for superconductivity in a multi-band system. 
As mentioned above, recent neutron scattering experiments\cite{neutron}
 have confirmed that local moments coupled with itinerant physics accounts 
 for the magnetism in the iron pnictides.

We first perform a Holstein-Primakoff transformation on the
spin-part of the Hamiltonian to obtain an effective description of
the spin excitations on the $(\pi,0)$ antiferromagnet.  To order
$1/S$, we
 obtain\cite{Yao},
\begin{eqnarray}
    H_{\rm s}&=&{\rm C_1}+S\sum [A_{\rm k}a^{\dag}_{{\bf k}}a_{{\bf k}}
    +\frac{1}{2}(B_{\bf k}a^{\dag}_{{\bf k}}a^{\dag}_{{\bf -k}}+B^*_{\bf -k}a_{{\bf k}}a_{{\bf -k}})]\nonumber\\
    A_{\bf k}&=&4J_2+2J_1\cos k_x\nonumber\\
    B_{\bf k}&=&4 J_1\cos k_y+8J_2\cos k_x \cos k_y
    %    E_{\rm cl}&=&-2J_2N S^2.
\end{eqnarray}
When diagonalized using $b_{\bf k}=\cosh\phi_{\bf k}a_{\bf
k}-\sinh\phi_{\bf k}a_{\bf -k}^{\dag}$ , this Hamiltonian yields a
dispersion for the magnons of the form
\begin{eqnarray}
    H_{\rm s}&=& {\rm C_2}+S \sum \omega_{\bf k} b_{\bf k}^{\dag}b_{\bf k}\\
    % H_{\rm s}&=& E_{\rm cl}+E_{\rm g}+S \sum \omega(\bf k) b_{\bf k}^{\dag}b_{\bf k}\\
    \omega({\bf k})&=&S \sqrt{ A_{\bf k}^2-B_{\bf k}^2 }\nonumber
    % E_{\rm g}&=&\frac{S}{2}\sum \left(-A_{\bf k}+\omega(\bf k)\right)
\end{eqnarray}
where ${\rm C_i} (i=1,2)$ are all constant. Although in our case
$S=1/2$ and hence strictly speaking the large spin expansion is
invalid, the Holstein-Primakoff transformation still retains the key
physics that the magnon is a massless boson.  Hence, we adopt this
appraoch here to investigagte superconductivity in the multi-band
pnictides.  It is now convenient to express the Fourier
transform of our original Hamiltonian in terms of these effective
spin excitations. By using the Fourier transform $c_{{\rm
l}\sigma}=\sqrt{2/N}\sum_{\rm k}e^{-i{\rm k}{\rm l}}\alpha_{{\rm
k}\sigma}$ and $c_{{\rm m}\sigma}=\sqrt{2/N}\sum_{\rm k}e^{-i{\rm
k}{\bf m}}\beta_{{\rm k}\sigma}$
%===================================================
which define  $\alpha_{{\rm k}\sigma}$ and $\beta_{{\rm k}\sigma}$
%===================================================
and of which $\bf l$ and $\bf m$ are the site indices of the spin up and spin down electrons on the
localized band respectively,
%===================================================
we can rewrite the Hamiltonian\cite{Yamada},
\begin{eqnarray}\label{ham}
    H&=&H_{\rm s}+H'_{\rm e}+H_{\rm sf}^{\rm xy}+H_{\rm sf}^{\rm z}\\
    H'_{\rm e}&=&\sum\sum_{\nu=\pm} E_{{\bf k}\sigma}^\nu d^{\dag\nu}_{{\bf k}\sigma} d_{{\bf k}\sigma}^\nu\nonumber\\
    H_{\rm sf}^{\rm xy}&=& -J\sqrt{2S}\sum_{\bf k,k',q}\sum_{\nu=\pm}\delta({\bf
    k'-k+q})\nonumber\\
    \times && \left[f_{\nu,\nu'}({\bf k, k'}) d^{\dag\nu'}_{{\bf k'}\uparrow} d_{{\bf
    k}\downarrow}^\nu a_{\bf q}^\dag +h.c.\right]\nonumber\\
    H_{\rm sf}^{\rm z} &=&J\sqrt{2S}\left[\sum\sigma c^{\dag}_{{\bf l}\sigma}c_{{\bf
    l}\sigma}\left(a^{\dag}_{{\bf l}}a_{{\bf l}}-\left<a^{\dag}_{{\bf l}}a_{{\bf
    l}}\right>\right)+(\bf l \longrightarrow \bf m)\right]\nonumber
\end{eqnarray}
in terms of the the spin normal modes, $a_{\bf k}$ and the
electronic degrees of freedom, $d^\pm_{\bf k\sigma}$, which are
linear combinations of $\alpha_{\bf k\sigma}$ and $\beta_{\bf
k\sigma}$. The parameters in this effective Hamiltonian are defined
by
\begin{eqnarray}\label{E}
    \Delta E &=& \left(\epsilon_{\bf k}-\epsilon_{\bf k+Q}\right)/2 \ ,\ E_{\bf k}=\sqrt{\left(\Delta E\right)^2+J^2}\ \ \ \ \ \ \ \ \ \ \ \ \ \\
      E_{\bf k \sigma}^{\rm \pm}&=&\left(\epsilon_{\bf k}+\epsilon_{\bf
    k+Q}\right)/2\pm E_{\bf k}.
  %  f_{\nu,\nu'}(\bf k, q, P)&=&f(\sin\theta_{\bf k-q-P},\sin\theta_{\bf k})\nonumber\\
  %  \sin\theta_{\bf k}&=& \sqrt{\frac{1}{2}\left(1-\frac{J}{E_{\bf
  %  k}}\right)} , \cos\theta_{\bf k}=\sqrt{\frac{1}{2}\left(1+\frac{J}{E_{\bf
  %  k}}\right)}\nonumber
\end{eqnarray}
where 
${\bf Q}=(\pi,0)$ in a wavevector for the magnetic order in the
unfolded Brillouin zone.   $f_{\nu,\nu'}(\bf k, k')$ is
given by\cite{Yamada,simple},
\begin{eqnarray}\label{f1}
  f_{++}(\bf k, k') &=&f_{--}(\bf k, k')=\sin(\theta_{\bf k'}-\theta_{\bf k}) 
  \end{eqnarray}
  \begin{eqnarray}\label{f2}
  f_{+-}(\bf k, k') &=&-f_{-+}(\bf k, k')=\cos(\theta_{\bf k'}-\theta_{\bf k}) \label{f2}
\end{eqnarray}
where $\sin\theta_{\bf k}= \sqrt{\left(1-J/E_{\bf k}\right)/2}$ and $\cos\theta_{\bf k}= \sqrt{\left(1+J/E_{\bf k}\right)/2}$. We can see from
these factors that intra-band scattering is suppressed when $\theta_{\bf k}\approx \theta_{\bf k'}$. This agrees with earlier work by
Schrieffer\cite{supression}. Fortunately, the inter-band scattering is
enhanced in this case which will play the key rule in the pairing mechanism.

\begin{figure}
  \includegraphics[width=3.5in]{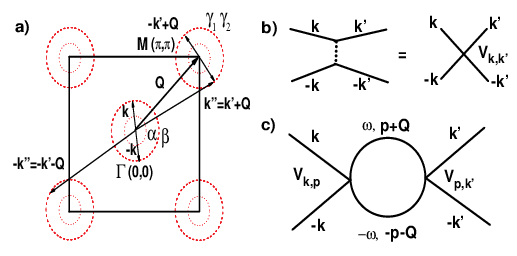}
  \caption{ a) Fermi surfaces and Cooper pairs (CP) of Iron-based SC on the folded Brillouin Zone.
   Two hole pockets ($\alpha$,$\beta$) on $\Gamma$ point
   and two electron pockets ($\gamma_1$,$\gamma_2$) at $M$ point are determined by dispersion relation
   $E^{\pm}_{\bf k}=\mu$ and $E^{\pm}_{\bf k+Q}=\mu$ respectively.
   Of two possible interband CP scattering process, $({\bf k}\uparrow, -{\bf k}\downarrow)$
scattered to $({\bf k^{''}}\downarrow, -{\bf k^{''}}\uparrow)$ and
$({\bf k}\uparrow, -{\bf k}\downarrow)$ scattered to $({\bf
k'+Q}\downarrow, {-\bf k'+Q}\uparrow)$, only in the first case is
momentum conserved. b) Interaction vertex mediated by magnons. c)
Attractive interaction mediated by interband scattering. }\label{FS}
\end{figure}

We emphasize that $\epsilon_{\bf k}$ is the bare spectrum of the
electron in the itinerant level. 
%======================================================
In the itinerant level, due to multiple orbital degree of freedom and the hopping to both nearest neighbor and
next nearest neighbor sites, the electron spectrum has a multi-band
structure. 
%======================================================
If we consider the  two-band tight
binding model\cite{2band}, we have the following electronic
spectrum,
\begin{eqnarray}
  \xi_{\pm}({\bf k}) &=& \epsilon_{+}({\bf k})\pm\sqrt{\epsilon^2_{-}({\bf k})+\epsilon^2_{xy}({\bf k})} \\
  \epsilon_{\pm}({\bf k}) &=&\frac{\epsilon_{x}({\bf k})+\epsilon_{y}({\bf k})}{2}  \nonumber\\
  \epsilon_{x}({\bf k}) &=& -2 t_1\cos k_x a-2 t_2\cos k_y a-4 t_3 \cos k_x a\cos k_y a \nonumber\\
   \epsilon_{y}({\bf k}) &=& -2 t_2\cos k_x a-2 t_1\cos k_y a-4 t_3 \cos k_x a\cos k_y a \nonumber\\
   \epsilon_{xy}({\bf k}) &=& -4 t_4 \sin k_x a\sin k_y a,\nonumber
\end{eqnarray}
where $t_1=-1$, $t_2=1.3$ and $t_3=t_4=-0.85$\cite{2band}. Fermi
surfaces in the folded Brillouin zone are shown in Fig.(\ref{FS}a).
Apparently, magnetic order will change the electronic spectrum. For
example, $\beta$ and $\gamma_2$ Fermi surfaces are determined by
$\xi_{-}({\bf k})=\mu$ and $\xi_{+}({\bf k})=\mu$ respectively if no
magnetic order. In the presence of antiferromagnetic order, they are
determined by $E_{\bf k}^{-}=\mu$ and $E_{\bf k}^{+}=\mu$ instead.

%Combined with Eq.(\ref{E}), in the parent compound and the
%superconducting state where long range magnetic order exist, we
%should choose $\epsilon_{\bf k}=\xi_{-}({\bf k})$ around $\Gamma$
%point $\epsilon_{\bf k}=\xi_{+}({\bf k})$ around M point. And for
%both case,  $E_{\bf k \sigma}^{\rm -}<\mu$ and $E_{\bf k
%\sigma}^{\rm +}>\mu$ determined electronic spectrum inside and
%outside Fermi surfaces respectively. So a gap in order of $J$ will
%developed in both electronic and hole Fermi surfaces. While long
%range magnetic order is destroyed, gap will disappear.

%\begin{eqnarray}\label{}
%&V({\bf q})&a^{\dag}_{\bf k_1+q,\uparrow}a^{\dag}_{\bf
%k_2-q,\downarrow}a_{\bf k_1,\downarrow}a_{\bf k_2\uparrow}\\
%=-&V({\bf q})&a^{\dag}_{\bf k_1+q,\uparrow}a^{\dag}_{\bf
%k_2-q,\downarrow}a_{\bf k_2,\uparrow}a_{\bf
%k_1\downarrow}\nonumber\\
%\left( \Delta^h_{\bf k} , \Delta^e_{\bf k}\right)^{\rm T}  &\propto&
%\left(\lambda_{\rm
%   eh},-\left(\lambda+\lambda_{\rm h}\right)\right)^{\rm T}\\
% V_{\nu,\nu'}&=& \frac{(2J^2S)|f_{\nu,\nu'}({\bf k, k',P})|^2}{\epsilon_e({\bf q})\omega({\bf
%    q})}\\
%   \epsilon_e({\bf q}\rightarrow 0)&\rightarrow & q_s/q+ 1\\
%    \epsilon_e({\bf q}\rightarrow {\bf Q})&\rightarrow & \rm const\\
%   \omega({\bf q}\rightarrow 0)&\rightarrow& q\ \rm const\\
%   \omega({\bf Q}+({\bf q}\rightarrow 0))&\rightarrow& q\ \rm const\\
%\end{eqnarray}

\subsection{Magnon-Mediated Interaction}
There is of course an extreme similarity between the various contributions to Eq. (\ref{ham})  and the
electron-phonon Hamiltonian used in BCS.  In this case, the phonons
are replaced by magnons and the interactions in $H^{\rm xy}_{\rm sf}$
involve spin flips.  It is from this term that the dominant
interactions arise. Treating this term perturbatively to second order, we obtai the amplitude
\begin{eqnarray}\label{Vee}
    V_{\nu,\nu'}&=&%\frac{V_e({\bf q})}{\epsilon_e({\bf q})}+
    -\frac{(2J^2S) \omega({\bf
    q})|f_{\nu,\nu'}({\bf k, k',P})|^2}{\left[\left(E_{\bf k'}^{\nu}-E_{\bf k}^{\nu}\right)^2-\omega^2({\bf
    q})\right]},
\end{eqnarray}
for the scattering of an electron pair with momenta $(\bf k,- k)$ to
one with momenta $(\bf k',-\bf k')$.  In Eq. (\ref{Vee}),
 $\bf q=\bf k'-\bf k$. Regardless of the momentum transfer, the interaction arising from magnon
scattering is always repulsive as a result of the minus sign in
front of Eq. (\ref{Vee}).  This crucial difference\cite{magnon}
with the phonon-mediated interaction in BCS arises from the
spin-flip nature of the magnon scattering.

Nonetheless,  superconductivity with different signs for the order parameter on the different Fermi surfaces is still possible even
though the scattering processes are repulsive.  To establish this we
consider the separate amplitudes for inter-band (between the $M$ and
$\Gamma$ points) and intra band (within the $M$ or $\Gamma$ points)
scattering processes. Intraband scattering dominates in the long
wavelength limit. Rewriting $q_x=q\cos\alpha$ and $q_y=q\sin\alpha$,
we have that $\omega({\bf q})=2S q
\sqrt{(2J_2+J_1)\left[(2J_2-J_1)\cos^2\alpha+(2J_2+J_1)\sin^2\alpha\right]}\equiv
2Sq F(\alpha)$ where $q=|{\bf q}|$.  For the intraband scattering,
$E_{\bf k'}^{\nu}-E_{\bf k}^{\nu}\approx 0$. Consequently, the
magnon-mediated interaction is of the form,
\begin{equation}\label{intra}
 V_{\rm ii}({\bf k, k'})=\frac{(J^2) |f_{\nu,\nu'}({\bf k,
 k'})|^2}{ q
 F(\alpha)},
\end{equation}
where { the subscript} $\rm ii$ denotes $\rm ee$ for intra-band
scattering within an electron pocket and $\rm hh$ for scattering
within a hole pocket. We compare this amplitude to that for
inter-band scattering. The only interband scattering process that is
momentum conserving involves the scattering of an electron pair from
the hole pocket at $\Gamma$ to two different electron pockets such
that the final pair has momentum $({\bf k'+Q}\downarrow, {-\bf
k'-Q}\uparrow)$. For the interband scattering, we still have $E_{\bf
k'}^{\nu}-E_{\bf k}^{\nu}\approx 0$ but because the momentum carried
by the magnon is $\bf Q+\bf q$, quantitative differences,
\begin{equation}\label{intra}
 V_{\rm eh}({\bf k, k'})=\frac{(J^2) |f_{\nu,\nu'}({\bf k,
 k'})|^2}{ q G(\alpha )},
\end{equation}
arise with the intra-band scattering amplitude. Here,
$G(\alpha)=\sqrt{2J_2(2J_2+J_1)}$.  Note the presence of the
inverse $q$ dependence seems to make both the intra-band and
inter-band interaction very large at the small $q$ limit. However,
$f_{\mu\nu}({\bf k},{\bf k'})$ has a quite difference dependence on
$\bf q$. Due to Adler's theorem\cite{Adler}, the interaction
induced by an exchange of a Goldstone excitation (here, it is
magnon) is proportional to the transferred momentum. For intra-band scattering,
the transferred momentum is {\bf q}. So the interaction vanishes
 as $q$ approaches zero. But for inter-band interaction, the
transferred momentum is $\bf Q+q$.  This is nonzero even though q approaches
zero. 
%=================================================
These conclusions can also be derived from Eq.(\ref{f1},\ref{f2}): $\lim_{q\rightarrow 0} f_{++}({\bf k,k+q})=\lim_{q\rightarrow 0} f_{--}({\bf k,k+q})
\propto\lim_{q\rightarrow 0} q=0$
and   $\lim_{q\rightarrow 0} f_{+-}({\bf k,k+q+Q})=-\lim_{q\rightarrow 0} f_{-+}({\bf k,k+q+Q})=1$.
%=================================================
Since $F(\alpha)$ and $G(\alpha)$ are of the same order, we
can conclude that the inter-band interaction will dominate the
intra-band interaction for small momentum $|\bf q|$. This is the
first demonstration from the microscopics that inter-band scattering
dominates in a realistic model for the pnictides.
%===================================================
This results is in fact universal: due to the Adler theorem,  the inter-band scattering dominates the
 intra-band scattering at long wave length limit  as long as  both interactions are induced by 
 Goldstone boson and the transferred momentum of inter-band interaction
 is non-zero.  
 %===================================================

The main difference between our model and the spin-fermion model or
Anderson-lattice is that there exists a superexchange interaction
between the localized spins. As in spin-fermion models, a
spatially oscillating RKKY interaction between the localized spins
will be induced by the itinerant electrons which is on the order of
$(J^2)$\cite{RKKY}. However, if we assume $J\ll J_1$ and $J\ll J_2$, the
RKKY interaction can be neglected.
From Eq.(\ref{ham}), we can see that a spin up electron, $d_{{\bf
k}\uparrow}$, is always coupled to a spin down electron,
$d_{{\bf k+q}\downarrow}$, in the electron-magnon interaction term.
Hence, our work is consistent with the spin-singlet pairing seen in Knight shift
experiments\cite{Kawabata}.

\subsection{$s_{\pm}$-Pairing Symmetry Induced By Magnon-Mediated Interaction}
In the limit that the inter-band scattering dominates, the BCS gap
equations admit sign-reversing pairing entirely
 from repulsive interactions. This can be seen by solving the gap equations
\begin{eqnarray}\label{lhs}
    \lambda \left(
                 \begin{array}{c}
                   \Delta^h_{\bf k} \\
                   \Delta^e_{\bf k}\\
                 \end{array}
               \right) &=& -\sum_{\bf k'}\left< \left(
                 \begin{array}{cc}
                   V_{\rm hh} & V_{\rm eh} \\
                   V_{\rm eh} & V_{\rm ee} \\
                 \end{array}
               \right)({\bf k, k'})  \left(
                 \begin{array}{c}
                   \Delta^h_{\bf k'} \\
                   \Delta^e_{\bf k'}\\
                 \end{array}
               \right) \right>_{\rm FS},
\end{eqnarray}
for a 2-band system. A crucial difference with earlier work is the
minus sign on the right-hand side of Eq. (\ref{lhs}). If we assume
all Fermi surfaces are circular, then integrating over the Fermi
surface is equivalent to multiplying by the DOS. Defining the
dimensionless interaction strength $\lambda_{\rm e}=V_{\rm ee}N_{\rm
e}(E_{\rm F})$, $\lambda_{\rm h}=V_{\rm hh}N_{\rm h}(E_{\rm F})$ and
$\lambda_{\rm eh}=V_{\rm eh}\sqrt{N_{\rm e}(E_{\rm F}) N_{\rm
h}(E_{\rm F})}$, we obtain the maximum positive eigenvalue,
\begin{eqnarray}\label{OP}
  \lambda &=& -\left(\lambda_{\rm h}+\lambda_{\rm e}\right)/2+ \sqrt{\left(\lambda_{\rm
  h}
  -\lambda_{\rm e}\right)^2/4+\lambda^2_{\rm eh}},
\end{eqnarray}
and the corresponding eigenstate $\left( \Delta^h_{\bf k} ,
\Delta^e_{\bf k}\right)^{\rm T}  \propto  \left(\lambda_{\rm
   eh},-\left(\lambda+\lambda_{\rm h}\right)\right)^{\rm T}$.
We see clearly that for $\lambda>0$, the order parameter on the hole
and electron Fermi surfaces has different signs only if
$\lambda_{\rm eh}$ is positive.  
%====================================================
This is the well-know results in iron-pnictides\cite{Singh,jph}.  
It is the multi-band structure of the pnictides  and the repulsive inter-band interaction that lead naturally to $s_{\pm}$ pairing symmetry.
%It is the multi-band structure of the pnictides  and the Goldstone nature
%of the glue boson (the Adler theorm) that make inter-band interaction
%dominate the intra-band interaction thus making superconductivity possible.
%But the sign-reversing superconductivity is due to the repulsive inter-band 
%interaction. So  $s_{\pm}$ pairing symmetry is a combining results of
%multi-band properties and repulsive interaction.
 %====================================================
The critical temperature is given by
\begin{equation}\label{}
    T_c=1.14\Omega_{\rm sf}\exp(-1/\lambda),
\end{equation}
where $\Omega_{\rm sf}$ is the counterpart of the Debye frequency
for magnons, which is estimated to be $1500\rm K$ for
$J_1=J_2=500\rm K$\cite{DFA}. Coupled with a large $\lambda$ as a
result of the dominance of $\lambda_{\rm eh}$, the transition
temperature can be sizeable.

 A simple physical argument is instructive here to delineate why pairing can
 obtain from repulsive inter-band interactions. Consider the second-order process
shown in Fig. (\ref{FS}c).  A Cooper pair at a hole pocket is scattered into
a Cooper pair at an
electron pocket and then scatted back to the hole pocket. Regardless of the
sign of the interband scattering, the resultant matrix element for this process
\begin{eqnarray}
  U({\bf k, k'}) &=& \int d\omega\frac{\int d^3 {\bf p} V_{\rm eh}({\bf k,p})V_{\rm eh}({\bf p, k'}) }{\left(\omega+i\delta-\epsilon_{\bf p+Q}\right)
  \left(-\Omega-\omega+i\delta-\epsilon_{\bf -p-Q}\right)} \nonumber\\
   &=& -\int d^3 {\bf p} \frac{ V_{\rm eh}({\bf k,p})V_{\rm eh}({\bf p, k'}) }{\Omega+\epsilon_{\bf p+Q}+\epsilon_{\bf
   -p-Q}}<0,
\end{eqnarray}
is always attractive\cite{Kondo,Tony}.  Here,
$-\Omega=-(\epsilon_{\bf k}+\epsilon_{\bf -k})$ is the energy of two
electrons on the hole pockets. The key point here is that as long as
electron and hole pockets are separated in energy, an attraction
develops through an exchange of two electrons rather than
single-electron hopping.

A similar type of mechanism has been proposed by
 A. Overhauser and J. Appel\cite{2magnon} in which the exchange of two magnons leads to an attractive
interaction between two electrons. The problem with this mechanism
 is that if the electron-magnon interaction is weak, exchanging one magnon will
induce repulsive interactions which will dominate the attractive interaction induced 
by the second order process.  But for iron-based superconductors, the situation
 is different due to the presence of multi-bands. A process in which
pairs of electrons hop to another Fermi surface and back to the
original Fermi surface always involves an even number of inter-band scattering process
which always results in an attraction.   In this case, if the inter-band process 
dominate the intra-band process which could be repulsive or attractive,
 the overall net interaction could be attractive. By solving the gap equation, 
 we in fact have sumed over all even orders of the inter-band scattering
processes.

A critique of magnon-mediated pairing, or more general, spin-fluctuation
mediated pairing, comes from two fronts. The first one was advanced by
Schrieffer\cite{supression}. The pairing potential due to exchanging one spin
fluctuation can be written as a product of two vertices $J^2\Gamma_{\bf k,q}^2$
and a magnetic susceptibility $\chi_{\bf q}$. Because both $\Gamma_{\bf k,q}=
\left[({\bf q-Q})^2+L_s^{-2}\right]^{1/2}$ and $\chi_{\bf q}\approx 1/[({\bf
q-Q})^2+i\omega]$ exhibit a dominant $(\bf q-Q)^2$ dependence, this effective
interaction is suppressed resulting in a flat, featureless pairing potential.
This is still true in our approach for the first-order intra-band scattering.
Fortunately, inter-band scattering which plays the key rule in the pairing
mechanism is enhanced. Further, as mentioned above, the lowest order involving
the inter-band process obtains at second order, which is similar to the
spin-bag pairing mechanism\cite{supression,spinbag}. The second critique
concerns whether long-range magnetic order can coexist with superconductivity.
In fact, short-ranged  magnetic order is sufficient to excite magnons. In our
itinerant-localized model, if doping only removes electrons from the itinerant
level, thereby making it possible for electrons to tunnel to it from the
localized level, superconductivity can be created by pair formation through
magnon-mediated inter-band scattering. This mechanism also weakens
magnetism\cite{Spalek,Takashi} but a threshold doping level must be exceeded
before true long-range magnetism ceases.  Consequently, there is a finite range
of doping where both superconductivity and magnetism could co-exist.  Thus,
these two orders can coexist at a microscopic level or in a phase separated
fashion around the phase boundary\cite{Coex, Coex1,Coex2}.  Their coexistence
should be more easily to observe in the 122 materials than in the 1111 systems
in which superconductivity can be induced by high pressure instead of doping.

In essence, interband scattering can be viewed as an internal Josephson junction with coupling $V_{\rm eh}\cos(\phi_1-\phi_2)$ where $\phi_1$ and
$\phi_2$ are the phases of the order parameter on the two Fermi surfaces.
Thus, a positive $V_{\rm eh}$ will automatically bring a $\pi$ phase
shift on the order parameters of the two Fermi surface.
A striking feature of the pnictides is the relative insensitivity of
$T_c$ to electronic doping.  Within our model, this is naturally explained
because the key factors that determine $T_c$ are the magnon energy and the coupling $\rm J$.  Since $J$ is simply determined by the localized
carriers, doping has a relatively weak effect if it doesn't completely destroy the local magnetic order. Predominantly, doping changes the density of
itinerant electrons as is confirmed experimentally\cite{dope}.

%Within our model, the interaction on the itinerant level is
%weak\cite{WPA} compared with the localized level, by increasing the
%pressure, the bandwidth will increase to exceed the interaction
%strength on itinerant level thus be able to create enough itinerant
%electrons to support a SC state. This also agree with the
%experiments on $\rm CaFe_2As_2$
%  that by only increasing pressure without any doping, SC state also
%  appear\cite{Presure}.

\section{Isotope Effect}
What about the isotope effect? In the phonon-mediated case, the
isotope effect appears through the dependence on the Debye
frequency. Because the counterpart of the Debye frequency,
$\Omega_{\rm sf}$, for spin-fluctuation pairing is not related to
the ion mass, there should be no such effect.  However, beside
changing the Debye frequency,  isotopic substitution also modifies
the lattice constant $\bf a$\cite{Schrieffer}.  Smaller values of
$\bf a$ will increase the electron-electron interaction, thereby
enhancing the spin fluctuations and, as a consequence, a higher
$T_c$. Experimentally, the critical temperature of the 122 materials
is sensitive to pressure\cite{Canfield} and hence consistent with
this explanation.  Experimentally, $T_c$ and $T_{\rm SDW}$ decrease
at a rate of $\Delta T_c/\Delta P\approx 0.22\rm K/kbar$ for $\rm
(Ba_{0.55}K_{0.45})Fe_2 As_2$ and $\Delta T_{\rm SDW}/\Delta
P\approx 1.0\rm K/kbar$ for $\rm Ba Fe_2 As_2$\cite{Canfield}.
Consequently, it is reasonable to consider a harmonic toy model in
which the isotopic effect arises from a modulation in the lattice
constant.  Within such a heuristic model\cite{zero}, the isotopic
parameter is given by
\begin{equation}\label{alpha}
    \alpha=-\frac{\Delta T_c/T_c}{\Delta
    M/M}=\frac{1}{2}\left[\frac{\sinh(x)}{x}-1\right]>0,
\end{equation}
where $x=\hbar\omega/k_B T$.
For $\rm Ba_{0.6}K_{0.4}Fe_2 As_2$, $T_c\approx 36\rm K$ and
$\alpha=0.4$\cite{isotope}.  Explaining this shift requires a
zero-point energy of iron of roughly $6.2 \rm meV$.  For the 122
materials, the zero-point energy of iron is not known.  However, for
iron metal,
 nuclear resonant inelastic x-ray scattering place the zero-point energy at $6\rm meV$\cite{zeroFe}.
 Hence, our estimate used here to obtain the isotope shift is reasonable.  This kind of estimate
 is also applicable to the cuprates in which a positive isotope effect is seen experimentally\cite{QL}.

\section{Conclusion}
Based on an itinerant-localized model of iron-pnictide superconductors, we calculated the magnon-mediated interaction and found that both the intra-band and
inter-band interactions are repulsive. However, the inter-band interaction dominates the intra-band interaction. 
%=======================================
We found this results is universal: due to the Adler theorem,  the inter-band scattering dominates the
 intra-band scattering at long wave length limit as long as  both interactions are induced by Goldstone boson
 and the transferred momentum of inter-band interaction is non-zero. 
%=======================================
Such properties of interaction lead naturally to well-known $s_{\pm}$ pairing symmetry even though the interactions are repulsive. 
The competition and coexistence of magnetic order and superconductivity were discussed.
Doping will destroy long range antiferromagnetic order but short-range magnetic order can still survive which is sufficient to support such a
magnon-mediated pairing mechanism. 
%=======================================
Our model is also consistent with the recently discovered coexistence of superconductivity and magnetic order in iron-pnictides.
%======================================= 
The isotope effect is due to the pressure effect caused by isotopic substitution. In a heuristic model, the zero
point energy of an iron atom is estimated to be $6.2\rm meV$.

\ack We thank A. Bernevig and I. Mazin for helpful conversations
 and the NSF DMR-0605769 for partial funding
of this work.


\section*{References}
\begin{thebibliography}{}

\bibitem{JACS} Kamihar Y {\it et al} 2008 {\it J. Am. Chem. Soc. } {\bf 130} 3296
%\bibitem{JACS} Y. Kamihara,T. Watanabe, M. Hirano and H. Hosono, J. Am. Chem. Soc. {\bf 130}, 3296 (2008).
\bibitem{Wang} Wang C, {\it et al} 2008 {\it Europhys. Lett.} {\bf 83} 67006
\bibitem{Cruze} Cruz C de la {\it et al} 2008 {\it Nature (London)} {\bf 453} 899 % magnetic order vs SC in La(O,F)FeAs
\bibitem{Boeri} Boeri L et al. 2008 {\it Phys. Rev. Lett.} {\bf 101} 026403 % not-phonon-meidate theory
\bibitem{PDOS} Qiu Y  {\it et al } 2008 {\it Phys. Rev. B} {\bf 78} 052508
\nonum Christianson A D {\it et al } 2008 {\it Phys. Rev. Lett.} {\bf 101}
157004
%A. D. Christianson, {\it et al } arXiv:0807.3370 ;
\nonum Higashitaniguchi S  {\it et al } 2008 {\it Phys. Rev. } B {\bf 78}
174507
%S. Higashitaniguchi, {\it et al } arXiv: 3968.0807 ;
\nonum Phelan D  {\it et al } {\it Phys. Rev.} B {\bf 79} 014519
%D. Phelan, {\it et al } arXiv: 0811.1215 % not-phone-mediate experiments

\bibitem{Singh} Mazin I I {\it et al } 2008 {\it Phys.
Rev. Lett.} {\bf 101} 057003  \nonum Cvetkovic V {\it et al } 2009 {\it
Europhys. Lett.} {\bf 85} 37002
\bibitem{jph} Seo K,  Bernevig A and Hu J P 2008 {\it
Phys. Rev. Lett.} {\bf 101} 206404
\bibitem{s} Parker D {\it et al } 2008 {\it Phys. Rev.} B {\bf 78} 134524
\nonum Chubukov A V {\it et al } 2008 {\it Phys. Rev. } B {\bf 78} 134512 \nonum %arXiv: 0807.3735 ;
Lee P A and Wen X G 2008 {\it Phys. Rev} B {\bf 78} 144517 \nonum  Maier T A
{\it et al } 2008 {\it Phys. Rev. } B {\bf 78} 020514 \nonum  Wang F {\it et al } 2009 {\it Phys. Rev. Lett.} {\bf 102} 047005 % arXiv:0807.0498. % s-wave theory

\bibitem{isotope} Liu R H {\it et al } {\it Preprint } arXiv: 0810.2694 % Isotope effect on 122
%\bibitem{xray} C. J. Zhang, {\it et al } arXiv: 0811.3268  % x-ray aborption, EPI is strong
%\bibitem{Bhoi} D. Bhoi, {\it et al } arXiv: 0807.3833 % Normal state properties of 1111 SC
%\bibitem{Nakai} Y. Nakai, {\it et al } arXiv:0808.2293 % Spin-fluctuation is senstive to doping
\bibitem{Schrieffer} Scalapino D J, Schrieffer J R and Wilkins J W 1966 {\it Phys. Rev.} {\bf 148} 263




\bibitem{Goko} Goko T {\it et al } 2009 {\it Phys. Rev. } B {\bf 80} 024508   % coexistence of static AF and SF in phase separation way
\bibitem{Coex} Gonnelli R S {\it et al } 2009 {\it Phys. Rev. } B.{\bf 79}
184526 \nonum  Felner I {\it et al } {\it Preprint } arXiv: 0805.2794 % AF and SC coexistence in macroscopic level

\bibitem{Coex1} Pratt D K {\it et al } 2009 {\it Phys. Rev. Lett. } {\bf 103} 087001(2009)
\nonum Christianson A D {\it et al } 2009 {\it Phys. Rev. Lett. } {\bf 103},
087002 \nonum  Condron C L {\it et al } 2009 {\it Phys. Rev. } B {\bf 79},
144523

\bibitem{Coex2}  Zhang Y {\it et al } 2009 {\it Phys. Rev. Lett. } {\bf 102} 127003
\nonum Drew A J {\it et al } 2009 {\it Nature Mater.} {\bf 8} 310



\bibitem{WPA} Wu J, Phillips P and Castro Neto A H 2008 {\it Phys. Rev.
Lett.} {\bf 101} 126401.

\bibitem{Kondo} Kondo J 1963 {\it Theo. Prog. of Phys.} {\bf 36} 1
\bibitem{zeroFe} Mao H K {\it et al } 2008 {\it Science} {\bf 292} 914


\bibitem{hassan} Hsieh D {\it et al } {\it Preprint} arXiv:0812.2289
\bibitem{si} Si Q and Abrahams E 2008 {\it Phys. Rev. Lett. } {\bf 101} 076401

\bibitem{sachdev} Xu C, Muller M and Sachdev S 2008 {\it Phys. Rev. } B {\bf 78} 020501

%\bibitem{jph} K. Seo, B. A. Bernevig and J. P. Hu, {\it Phys. Rev. Lett. } {\bf 101}, 206404 (2008).

\bibitem{Dong} Dong J 2008 {\it et al } {\it Europhys. Lett.} {\bf 83} 27006
\bibitem{McGuire} McGuire M A {\it et al } 2008 {\it Phys. Rev. } B {\bf 78} 094517
\bibitem{Vild}  McGuire M A {\it et al } 2008 {\it Phys. Rev. } B {\bf 78} 064518
\bibitem{local} Wu T {\it et al } 2009 {\it Phys. Rev. } B {\bf 79} 115121
%arXiv:0811.2567.

\bibitem{hybridization} Garcia D R {\it et al } 2008 {\it Phys. Rev. } B {\bf 78} 245119  %arXiv: 0810.3034 %
%\bibitem{Garcia} M. A. McGuire, {\it et al } {\it Phys. Rev. } B {\bf 78}, 094517 (2008) % arXiv:0806.3878. z-magnetic moment
\bibitem{neutron} Zhao J {\it et al } {\it Nat. Phys.} published online: 13 July 2009 DOI 10.1038/nphys1336.



\bibitem{magnon} Vonsovskii S V {\it et al } 1966 {\it Sov. Phys. Usp. } {\bf 78} 723
\nonum Allen P B {\it et al } 1982 {\it Phys. Sol. Stat.} {\bf 37} 1
\bibitem{Yamada} Yamada H and Takada S 1974 {\it Prog. Theor. Phys.} {\bf 52} 1077  % s-f model
\bibitem{DFA}  Yildirim T 2008  {\it Phys. Rev. Lett. } {\bf 101} 057010

\bibitem{sawatzky} Sawatzky G A, Elfimov I S, van den Brink J and Zaanen J  {\it Preprint} arXiv: 0808.1390.

\bibitem{Yao} Yao D X and Carlson E W 2008 {\it Phys. Rev. } B {\bf 78} 052507
%arxiv: 0804.4115. % spin-wave excitation


\bibitem{simple}Our calculation is similar to that in the original paper\cite{Yamada},
but we have made several modifications: 1) We use the normal Fourier
transformation instead of $c_{{\rm m}\sigma}=\sqrt{2/N}\sum_{\rm
k}e^{-i{\rm k}{\bf (m+a)}}\beta_{{\rm k}\sigma}$ with ${\bf
a}=(a,0)$; 2) We neglected phase factors $e^{i n \pi}$ and $e^{i \bf
P a}$ since they have no effect on the physical quantities; 3) We
use the same type of Holstein-Primakoff transformation\cite{Yao} for
$\bf l$ and $\bf m$ sites thus obtain one kind of magnon operator
instead of two; 3) Reciprocal vector $\bf P$ is neglected since it
doesn't have any physical consequence here.


\bibitem{supression} Schrieffer J R 1995 {\it J. Low Temp. Phys.} {\bf
99} 397
\bibitem{2band} Raghu S {\it et al } 2008 {\it Phys. Rev. } B {\bf 77} 220503


%\bibitem{Pines} D. Pines, Phys. Rev. Vol.{\bf 109}, 280 (1958). % dielectric constant
%\bibitem{unscreen} Yuki Fuseya, {\it et al } arXiv: cond-mat/0811.3052. % unscreen effect

\bibitem{Adler} Adler S L 1965 {\it Phys. Rev.} {\bf 137} B 1022 \nonum Adler S L 1965 {\it Phys. Rev.} {\bf 139} B
1638 \nonum  Bourque A {\it et al } {\it Preprint } hep-ph/0509038.

\bibitem{RKKY} Ruderman M A and Kittel C 1954 {\it Phys. Rev.} {\bf
96}  99

\bibitem{Kawabata} Kawabata A, Lee S C, Moyoshi T,  Kobayashi Y and Sato M {\it Preprint} cond-mat/0807.3480 % knight shift: spin-single

\bibitem{Tony} Leggett A J 1966 {\it J. Theor. Phys.} {\bf 36} 901

\bibitem{2magnon}  Appel J and Overhauser A W 1994 {\it Physica} B {\bf
199 \& 200} 310



\bibitem{spinbag} Schrieffer J R, Wen X G, Zhang S C 1988 {\it Physica
Scripta} {\bf T27} 99




\bibitem{Spalek} Spalek J 1988 {\it Phys. Rev. } B {\bf 38} 208
\nonum Byczuk K {\it et al } 1992 {\it Phys. Rev. } B {\bf 46} 14134
\bibitem{Takashi} Yanagisawa T, Miyazaki M and Yamaji K 2009 {\it J. Phys. Soc. Jap.} {\bf 78} 013706


%\bibitem{Weng} S. P. Kou, {\it et al } arXiv: 0811.4111. % coupled local moments and itinerant electrons in Iron-based SC




\bibitem{dope} Hess C  {\it et al } {\it Preprint} arxiv: 0811.1601


%\bibitem{Zocco} D. A. Zocco, {\it et al } arXiv: 0805.4372.
 % Tc as non monotonic function of pressure
\bibitem{Canfield} Torikachvili M S, Bud'ko S L, Ni N, Canfield P C  {\it Preprint} arXiv: 0809.1080

\bibitem{zero} Covington E J and Montgomery D J 1957 {\it J. Chem. Phys.} {\bf 27} 1030  % lattice constant and isotope atoms

\bibitem{QL} Leggett A J 2006 {\it  Quantum Liquids} (Oxford: Oxford University).


\end{thebibliography}
\end{document}